\documentstyle[prd,aps,epsf]{revtex}

\def\simgt{\,\rlap{\lower 3.5pt\hbox{$\mathchar\sim$}}\raise 1pt\hbox {$>$}\,}
\def\simlt{\,\rlap{\lower 3.5pt\hbox{$\mathchar\sim$}}\raise 1pt\hbox {$<$}\,}

\newcommand{\bra}{\left\langle}
\newcommand{\ket}{\right\rangle}

\newcommand{\real}{\mbox{Re}\ }
\newcommand{\trace}{\mbox{Tr}\ }

\begin{document}

\draft

\tightenlines

\title{
{\normalsize \hfill {\sf UTHEP-443}} \\
\vspace*{-2pt}
{\normalsize \hfill {\sf UTCCP-P-105}} \\
\vspace*{-2pt}
{\normalsize \hfill {\sf May, 2001}} \\
Thermodynamics of SU(3) gauge theory on anisotropic lattices
}

\author{CP-PACS Collaboration :
  Y.~Namekawa\rlap,$^{\rm a}$
  S.~Aoki\rlap,$^{\rm a}$
  R.~Burkhalter\rlap,$^{\rm a,b}$ 
  S.~Ejiri\rlap,$^{\rm b}$ 
  M.~Fukugita\rlap,$^{\rm c}$
  S.~Hashimoto\rlap,$^{\rm d}$
  N.~Ishizuka\rlap,$^{\rm b}$
  Y.~Iwasaki\rlap,$^{\rm a,b}$
  K.~Kanaya\rlap,$^{\rm a}$ 
  T.~Kaneko\rlap,$^{\rm d}$ 
  Y.~Kuramashi\rlap,$^{\rm d}$
  V.~Lesk\rlap,$^{\rm b}$
  M.~Okamoto\rlap,$^{\rm b}$
  M.~Okawa\rlap,$^{\rm d}$ 
  Y.~Taniguchi\rlap,$^{\rm a}$
  A.~Ukawa$^{\rm b}$ and
  T.~Yoshi\'e$^{\rm b}$
 }

\address{
$^{\rm a}$Institute of Physics,
    University of Tsukuba, Tsukuba, Ibaraki 305-8571, Japan \\
$^{\rm b}$Center for Computational Physics,
    University of Tsukuba, Tsukuba, Ibaraki 305-8577, Japan \\
$^{\rm c}$Institute for Cosmic Ray Research,
    University of Tokyo, Kashiwa 277-8582, Japan \\
$^{\rm d}$High Energy Accelerator Research Organization
    (KEK), Tsukuba, Ibaraki 305-0801, Japan
}

\date{\today}
\maketitle


\begin{abstract}
Finite temperature SU(3) gauge theory is studied 
on anisotropic lattices using the standard plaquette gauge action. 
The equation of state is calculated on 
$16^{3} \times 8$, $20^{3} \times 10$ and $24^{3} \times 12$ 
lattices with the anisotropy $\xi \equiv a_s / a_t = 2$, 
where $a_s$ and $a_t$ are the spatial and temporal lattice 
spacings. 
Unlike the case of the isotropic lattice on which $N_t=4$ data deviate 
significantly from the leading scaling behavior,
the pressure and energy density 
on an anisotropic lattice are found to satisfy well the leading $1/N_t^2$ 
scaling from our coarsest lattice, $N_t/\xi=4$.
With three data points at $N_t/\xi=4$, 5 and 6, 
we perform a well controlled continuum extrapolation of
the equation of state. 
Our results in the continuum limit agree with a previous 
result from isotropic lattices using the same action,
but have smaller and more reliable errors. 
\end{abstract}


\section{Introduction}
\label{sec:intro}

Study of lattice QCD at finite temperatures
is an important step towards clarification of the dynamics of
the quark gluon plasma which is believed to have formed 
in the early Universe and is expected to be created 
in high energy heavy ion collisions \cite{ejiri}. 
In order to extract predictions for the real world from 
results obtained on finite lattices, we have to extrapolate 
lattice data to the continuum limit of vanishing lattice spacings.
Because of the large computational demands for full QCD simulations,
continuum extrapolations of thermodynamic quantities have so far been 
attempted only in SU(3) gauge theory, 
{\it i.e.}, in the quenched approximation of QCD, 
where the influence of dynamical quarks is neglected.
Two studies using the standard plaquette gauge action \cite{boyd} 
and a renormalization-group (RG) improved gauge action \cite{okamoto}
have found the pressure and energy density consistent with each other 
in the continuum limit.

In full QCD with two flavors of dynamical quarks, thermodynamic 
quantities on coarse lattices have been found to show large lattice spacing 
dependence \cite{milc,bielefeld,cppacs_ft}. 
For a reliable extrapolation to the continuum limit, data on finer 
lattices are required. 
With conventional isotropic lattices, this means an 
increase of the spatial lattice size to 
keep the physical volume close to the thermodynamic limit.
Full QCD simulations on large lattices are still difficult with the current 
computer power. 
A more efficient method of calculation is desirable.
Even in the quenched case, we note that continuum extrapolations of 
equation of state have been 
made using only two lattice spacings \cite{boyd,okamoto}.
In order to reliably estimate systematic errors from the 
extrapolations, more data points are needed. 
Therefore, an efficient method is called for also in quenched QCD. 

Recently, anisotropic lattices have been employed
to study transport coefficients and temporal correlation functions 
in finite temperature QCD \cite{sakai,taro,umeda}. 
In these studies, anisotropy was introduced to obtain more data points 
for temporal correlation functions.

In this paper, we show that anisotropic lattices provide also
an efficient calculation method for thermodynamic quantities. 
The idea is as follows.
Inspecting the free energy density of SU(3) gauge theory 
in the high temperature Stephan-Boltzmann limit, the
leading discretization error from the temporal direction is found to be 
much larger than that from each of the spatial directions. 
Hence, choosing $\xi=a_s/a_t$ larger than one, where $a_s$ and $a_t$
are the spatial and temporal lattice spacings, cutoff errors in 
thermodynamic quantities will be efficiently reduced without much 
increase in the computational cost. 
From a study of free energy density in the high temperature limit, 
we find that $\xi=2$ is an optimal choice for SU(3) gauge theory.
This improvement also makes it computationally easier to accumulate 
data for more values of temporal lattice sizes for the continuum 
extrapolation. 

As a first test of the method, we study the equation 
of state (EOS) in SU(3) gauge theory.
On isotropic lattices, discretization errors in the EOS for the plaquette
action are quite large at the temporal lattice size $N_t=4$.
The data at this value of $N_t$ deviate significantly  
from the leading $1/N_t^2$ scaling behavior,
$
\left. F(T) \right|_{N_t} 
= \left. F(T)\right|_{\rm continuum} + c_{F}/N_t^{2}
$, 
where $F$ is a thermodynamic quantity at a fixed temperature $T$. 
So far, continuum extrapolations of the EOS have been made using results at 
$N_t=6$ and 8.
On anisotropic lattices with $\xi=2$, 
we find that the discretization errors in the pressure and energy 
density are much reduced relative to those from 
isotropic lattices with the same spatial lattice spacing.
Furthermore, we find that the EOS at $N_t/\xi=4$, 5 and 6 follow 
the leading $1/N_t^2$ scaling behavior remarkably well.
Therefore, a continuum extrapolation can be reliably carried 
out. 
Since the total computational cost is still lower than that 
for an $N_t=8$ isotropic simulation, we can achieve a higher 
statistics as well, resulting in smaller final errors.

In Sec.~\ref{sec:highT_limit}, we study 
the high temperature limit of SU(3) gauge theory on anisotropic 
lattices to see how $\xi$ appears in the leading discretization 
error for the EOS. From this study, we find that $\xi=2$ is an optimum 
choice for our purpose.
We then perform a series of simulations on $\xi=2$ 
anisotropic lattices. 
Our lattice action and simulation parameters are described in 
Sec.~\ref{sec:simulation}. 
Sec.~\ref{sec:scale} is devoted to a calculation of the lattice scale 
through the string tension. 
The critical temperature is determined in Sec.~\ref{sec:Tc}.
Our main results are presented in Secs.~\ref{sec:pressure} and 
\ref{sec:energy}, where the pressure and energy density are calculated
and their continuum extrapolations are carried out. 
A brief summary is given in Sec.~\ref{sec:summary}.

\section{High temperature limit}
\label{sec:highT_limit}

In the high temperature limit, the gauge coupling vanishes due to 
asymptotic freedom, and SU(3) gauge theory turns into a free bosonic gas.
In the integral method \cite{engels4} which we apply in this study, 
the pressure $p$ is related to 
the free energy density $f$ by $p=-f$ for large homogeneous systems.
Therefore, in the high temperature limit, 
the energy density $\epsilon$ is given by 
$ 
 \epsilon = 3p = -3f.
$ 
The value of $f$ in the high temperature limit has been calculated in 
\cite{engels1,elze}. 
Normalizing $\epsilon$ by the Stephan-Boltzmann value in the continuum limit, 
we find 
\begin{equation}
 \frac{\epsilon}{\epsilon_{SB}} 
 = 1 + \frac{5 + 3 \xi^{2}}{21} 
       \left( \frac{\pi}{N_t} \right)^{2}
     + \frac{91 + 210 \xi^{2} + 99 \xi^{4}}{1680} 
       \left( \frac{\pi}{N_t} \right)^{4}
     + O\left( \left( \frac{\pi}{N_t} \right)^{6} \right)
\label{anisotropic-integral}
\end{equation}
for spatially large lattices. 
Substituting $\xi = 1$ in Eq.~(\ref{anisotropic-integral}), 
we recover the previous results for isotropic lattices \cite{scheideler}.
When we alternatively adopt the derivative method (operator method) 
\cite{engels1} to define the energy density, we obtain
\begin{equation}
 \frac{\epsilon}{\epsilon_{SB}}
 = 1 + \frac{5(1 + \xi^2)}{21} 
       \left( \frac{\pi}{N_t} \right)^{2}
     + \frac{13 + 50 \xi^{2} + 33 \xi^{4}}{240} 
       \left( \frac{\pi}{N_t} \right)^{4}
     + O\left( \left( \frac{\pi}{N_t} \right)^{6} \right).
\label{anisotropic-derivative}
\end{equation}
In both formulae, the leading discretization error is proportional 
to $1/N_t^2$. 

In the leading $1/N_t^2$ term of Eq.~(\ref{anisotropic-integral}) 
(or Eq.~(\ref{anisotropic-derivative})), 
the term proportional to $\xi^2$ represents the discretization error from 
finite lattice spacings $a_s$ in the three spatial directions. 
We find that the temporal cutoff $a_t$ leads to $5/8$ (or 1/2)
of the leading discretization error at $\xi=1$, 
while the spatial cutoff $a_s$ contributes only 1/8 (or 1/6) 
from each of the three spatial directions.

Since a reduction of the lattice spacing in each direction separately 
causes an increase of the computational cost by a similar magnitude, 
a reduction of $a_t$ is much more efficient than 
that of $a_s$ in suppressing lattice artifacts 
in thermodynamic quantities. 
Making the anisotropy $\xi = a_s/a_t$ too large is, however, 
again inefficient because the 
spatial discretization errors remain even in the limit of $\xi=\infty$. 
A rough estimate for the optimum value of $\xi$ is given by 
equating the discretization errors from spatial 
and temporal directions, 
$\xi = \sqrt{5} \approx 2.24$ from Eq.~(\ref{anisotropic-integral}),
and $\xi = \sqrt{3} \approx 1.73$ from Eq.~(\ref{anisotropic-derivative}). 
More elaborate estimations considering the balance between 
the computational cost as a function of the lattice size 
and the magnitude of discretization errors 
including higher orders of $1/N_t$ 
lead to similar values of $\xi$. 

Based on these considerations, we adopt $\xi = 2$ for simulations 
of SU(3) gauge theory in the present work. 
An even number for $\xi$ is attractive also for the 
vectorization/parallelization of the simulation code
which is based on an even-odd algorithm, since
we can study the case of odd $N_t/\xi$ without modifying the 
program. 

\section{Details of simulations}
\label{sec:simulation}

\subsection{Action}

We employ the plaquette gauge action for SU(3) gauge theory 
given by 
\begin{equation}
S_{G}[U] 
= \beta
  \left( 
    \frac{1}{\xi_0} Q_s
    + \xi_0 Q_t
  \right),
\label{lat-gauge-aniso}
\end{equation}
where $\xi_0$ is the bare anisotropy, $\beta=6/g_0^2$ with
$g_0$ the bare gauge coupling constant, 
and 
\begin{equation}
Q_s = \sum_{n,(ij)} \left(1 - P_{ij}(n)\right),\;\;
Q_t = \sum_{n,i} \left(1 - P_{i4}(n)\right),
\end{equation}
with $P_{\mu\nu}(n) = \frac{1}{3} \real \trace U_{\mu\nu}(n)$ 
the plaquette in the $(\mu,\nu)$ plane at site $n$. 
Anisotropy is introduced by choosing $\xi_0 \neq 1$. 

Due to quantum fluctuations, the actual anisotropy 
$\xi \equiv a_s / a_t$ deviates from the bare value $\xi_0$. 
We define the renormalization factor $\eta(\beta,\xi)$ for $\xi$ by 
\begin{equation}
 \eta(\beta,\xi) = \frac{\xi}{\xi_{0}(\beta,\xi)}.
\end{equation}
The values of $\eta(\beta,\xi)$ can be determined non-perturbatively
by matching Wilson loops in temporal and spatial directions
on anisotropic lattices \cite{scheideler,burgers,fujisaki,klassen}.
For our simulation, we calculate $\xi_0(\beta,\xi=2)$ 
using $\eta(\beta,\xi)$ obtained by Klassen 
for the range $1 \leq \xi \leq 6$ and $5.5 \leq \beta \leq \infty$
\cite{klassen}:
\begin{equation}
\eta(\beta,\xi) = 1 + \left( 1 - \frac{1}{\xi} \right) 
                      \frac{\hat{\eta}_{1}(\xi)}{6} \,
                      \frac{1+a_{1}g_{0}^{2}}{1+a_{0}g_{0}^{2}} g_{0}^{2},
\label{renormalize_anisotropy}
\end{equation}
where  $a_{0}= -0.77810$, $a_{1} = -0.55055$ and
\begin{equation}
\hat{\eta}_{1}(\xi) = \frac{1.002503 \xi^{3} + 0.39100 \xi^{2} 
                            + 1.47130 \xi - 0.19231}
            {\xi^{3} + 0.26287 \xi^{2} + 1.59008 \xi -0.18224}.
\end{equation}

\subsection{Simulation parameters}

The main runs of our simulations are carried out on $\xi = 2$ anisotropic 
lattices with size $N_s^3\times N_t = 16^{3} \times 8$, 
$20^{3} \times 10$ and $24^{3} \times 12$. 
For $N_t=8$, we make additional runs on $12^3\times8$ and $24^3\times8$ 
lattices to examine finite size effects. 
The zero-temperature runs are made on $N_s^3\times \xi N_s$ lattices 
with $\xi=2$. 
The simulation parameters of these runs which cover the range 
$T/T_c \sim 0.9$--5.0 are listed in Table~\ref{tab:simulation_parameters}.
To determine precise values for the critical coupling, longer runs around 
the critical points are made at the parameters compiled
in Table~\ref{tab:simulation_parameters-critical_coupling}.

For the main runs, 
the aspect ratio $L_sT = (N_s a_s)/(N_t a_t)$ is fixed to 4, 
where $L_s = N_s a_s$ is the spatial lattice size in physical units. 
This choice is based on a study of finite spatial 
volume effects presented in Sec.~\ref{sec:pressure}, 
where it is shown that, 
for the precision and the range of $T/T_c$ we study, 
finite spatial volume effects in the EOS are sufficiently small 
with $L_s T \geq 4$.

\begin{figure}[tb]
\begin{center}
\leavevmode
\epsfxsize=12cm\epsfbox{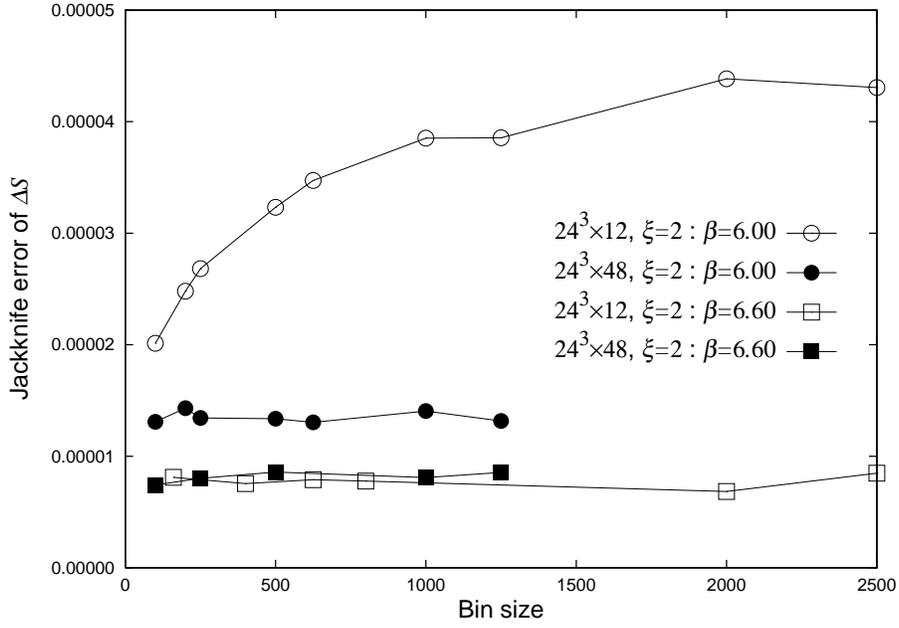}
\caption{Typical bin size dependence of jack-knife errors for $\Delta S$.}
\label{fig:jackknife}
\end{center}
\end{figure}

Gauge configurations are generated by a 5-hit pseudo heat bath update 
followed by four over-relaxation sweeps, which we call an iteration.
As discussed in Sec.~\ref{sec:pressure}, the total number of iterations 
should be approximately proportional to $N_t^{6}$ to keep an accuracy for the EOS.
After thermalization, 
we perform 20,000 to 100,000 iterations on finite-temperature lattices
and 5,000 to 25,000 iterations on zero-temperature lattices, 
as compiled in Table~\ref{tab:simulation_parameters}. 
At every iteration, we measure the spatial and temporal plaquettes, 
$P_{ss}$ and $P_{st}$. 
Near the critical temperature, we also measure the Polyakov loop.
The errors are estimated by a jack-knife method. 
The bin size for the jack-knife errors, 
listed in Table~\ref{tab:simulation_parameters}, 
is determined from a study of bin size dependence
as illustrated in Fig.~\ref{fig:jackknife}.
The results for the plaquettes are summarized 
in Tables~\ref{tab:MC_results-16x8_32}--\ref{tab:MC_results-24x12_48}.

\section{Scale}
\label{sec:scale}

\subsection{Static quark potential}
\label{subsec:pot}

We determine the physical scale of our lattices from the string tension, 
which is calculated from the static quark-antiquark potential 
at zero temperature. 
%
To calculate the static quark potential, 
we perform additional zero-temperature simulations listed in 
Table~\ref{tab:pot_simulations}. 
The static quark potential $V(\hat{R})$ is defined through 
\begin{equation}
 W(\hat{R},\hat{T}) 
 = C(\hat{R}) e^{-V(\hat{R}) \hat{T} / \xi},
\label{eq:potential}
\end{equation}
where $W(\hat{R},\hat{T})$ is the Wilson loop 
in a spatial-temporal plane 
with the size $\hat{R} a_s \times \hat{T} a_t$.
We measure Wilson loops at every 25 iterations after thermalization.
In order to enhance the ground state signal in (\ref{eq:potential}),
we smear the spatial links of the Wilson loop \cite{bali,bali2}. 
Details of the smearing method are the same as in Ref.\cite{cppacs_pot}.
We determine the optimum smearing step $N_{opt}$ which maximizes 
the overlap function $C(\hat{R})$ 
under the condition $C(\hat{R}) \le 1$.
Following Ref.\cite{bali2}, we study a local effective potential 
defined by
\begin{equation}
 V_{\it eff}(\hat{R},\hat{T}) 
 = \xi \log \left( \frac{W(\hat{R},\hat{T})}{W(\hat{R},\hat{T}+1)}
            \right),
\label{eq:potential2}
\end{equation}
which tends to $V(\hat{R})$ at sufficiently large $\hat{T}$. 
The reason to adopt Eq.~(\ref{eq:potential2}) instead of the fit result 
from Eq.~(\ref{eq:potential}) is to perform a correlated error analysis
directly for the potential parameters. 
The optimum value of $\hat{T}$, listed in Table~\ref{tab:string_beta}, 
is obtained by inspecting the plateau of $V_{\it eff}(\hat{R},\hat{T})$
at each $\beta$. 

We perform a correlated fit of 
$V(\hat{R}) = V_{\it eff}(\hat{R},\hat{T}_{opt})$ 
with the ansatz \cite{michael},
\begin{equation}
 V(\hat{R}) = V_{0} + \sigma \hat{R} - e \frac{1}{\hat{R}}
                       + l \left( \frac{1}{\hat{R}}
                       - \left[\frac{1}{\hat{R}}\right]  \right).
\label{pot-fit}
\end{equation}
Here, $\left[\frac{1}{\hat{R}}\right]$ is the lattice Coulomb 
term from one gluon exchange
\begin{equation}
 \left[\frac{1}{\hat{R}}\right] 
 = 4\pi \int_{-\pi}^{\pi} \frac{d^{3}\mathbf{k}}{(2\pi)^{3}}
                         \frac{\cos (\mathbf{k} \cdot \hat{R})}
                              {4 \sum_{i=1}^{3} \sin^{2} 
                               (k_{i}a_s/2)},
\end{equation}
which is introduced to approximately remove terms violating rotational 
invariance at short distances. 
The coefficient $l$ is treated as a free parameter. 

The fit range  $[\hat{R}_{min},\hat{R}_{max}]$ for $\hat{R}$
is determined by consulting the stability of the fit. 
Our choices for $\hat{R}_{min}$ are given in Table~\ref{tab:string_beta}.
We confirm that the fits and the values of the string tension are 
stable under a variation of $\hat{R}_{min}$.
The string tension is almost insensitive to a wide variation of 
$\hat{R}_{max}$. Hence
$\hat{R}_{max}$ is chosen as large as possible so far as the fit is stable 
and the signal is not lost in the noise.
With this choice for the fit range, we obtain fit curves 
which reproduce the data well. 

Our results for the potential parameters are summarized in 
Table~\ref{tab:string_beta}.
The error includes the jack-knife error with bin size one (25 iterations) 
and the systematic error from the choice of $\hat{R}_{min}$ 
estimated through a difference under the change of $\hat{R}_{min}$ by one. 
We confirm that 
increasing the bin size to two gives consistent results 
on $16^3\times 32$ lattices, while, on $24^3\times 48$ lattices, 
correlated fits with bin size two become unstable due to 
insufficient number of jackknife ensembles.

\subsection{String tension}
\label{subsec:sigma}

We interpolate the string tension data 
using an ansatz proposed by Allton \cite{allton}, 
\begin{equation}
 a_s \sqrt{\sigma} 
 = f(\beta) \, \frac{1 + c_{2}\hat{a}(\beta)^{2} 
                    + c_{4}\hat{a}(\beta)^{4}}
                 {c_{0}},
\label{allton_fitting_eq}
\end{equation}
where $f(\beta)$ is the two-loop scaling function of 
SU(3) gauge theory,
\begin{eqnarray}
  f(\beta) &=& \left(\frac{6b_{0}}{\beta}\right)^{- \frac{b_{1}}{2 b_{0}^{2}}}
                  \exp[-\frac{\beta}{12 b_0}], \nonumber \\
 b_{0} &=& \frac{11}{16 \pi^{2}}, \;\;
 b_{1} = \frac{102}{(16 \pi^{2})^{2}},
\label{allton_fitting_eq2}
\end{eqnarray}
and  $\hat{a}(\beta) \equiv f(\beta)/f(\beta=6.0)$.

\begin{figure}[tb]
\begin{center}
\leavevmode
\epsfxsize=12cm\epsfbox{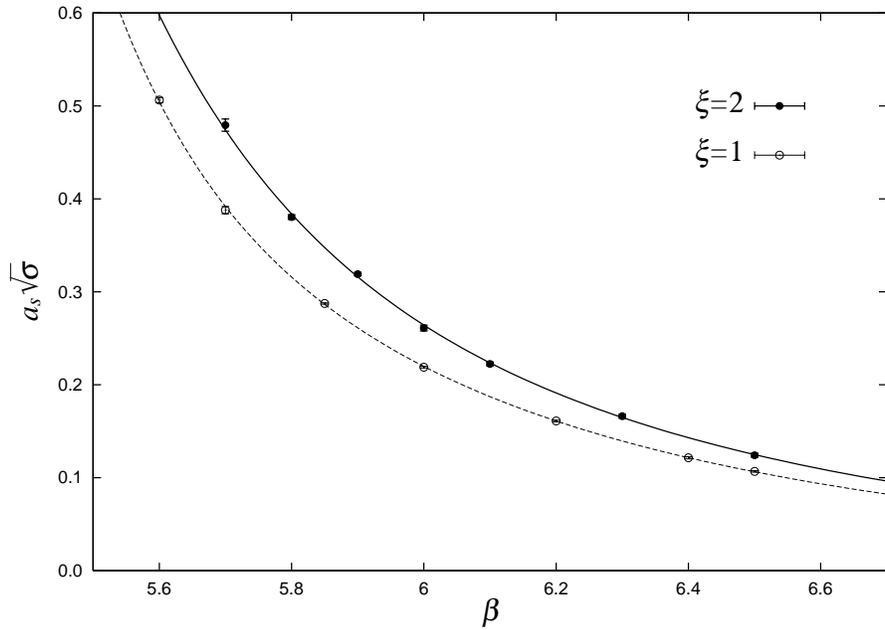}
\caption{String tension $\sigma$ on $\xi = 2$ anisotropic lattices
	as a function of $\beta$. Scaling fits are based on the ansatz 
	(\protect{\ref{allton_fitting_eq}}).
	}
\label{fig:allton-fit}
\end{center}
\end{figure}

From Table~\ref{tab:string_beta}, we find that the values for 
$a_s\sqrt{\sigma}$ are insensitive to the spatial lattice volume 
to the present precision. 
Using data marked by star ($*$) in Table~\ref{tab:string_beta}, 
we obtain the best fit at 
\begin{equation}
c_{0} = 0.01171(41), \;\; 
c_{2} = 0.285(79), \;\; 
c_{4} = 0.033(30),
\end{equation}
with $\chi^{2}/N_{DF} = 1.77$. 
The string tension data and the resulting fit curve are shown 
in Fig.~\ref{fig:allton-fit}, together with those from 
isotropic lattices \cite{edwards}.

\section{Critical temperature}
\label{sec:Tc}

We define the critical gauge coupling $\beta_{c}(N_t,N_s)$ from the 
location of the peak of the susceptibility $\chi_{rot}$ for a 
Z(3)-rotated Polyakov loop. 
The simulation parameters for the study of $\beta_{c}$ are compiled
in Table~\ref{tab:simulation_parameters-critical_coupling}.
The $\beta$-dependence of $\chi_{rot}$ is calculated 
using the spectral density method \cite{swendsen}. 
The results for $\beta_c$ are compiled in Table~\ref{tab:beta_c_lat}.

\begin{figure}[tb]
\begin{center}
\leavevmode
\epsfxsize=12cm\epsfbox{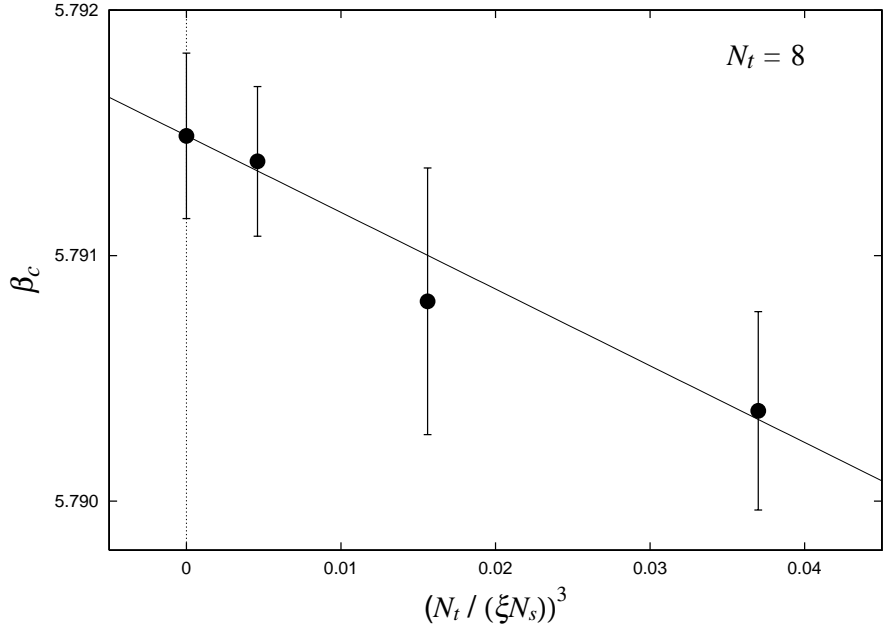}
\caption{Finite-size scaling of $\beta_c$ for $N_t/\xi=4$
         on $\xi = 2$ anisotropic lattices.
         }
\label{fig:finite_size_scaling}
\end{center}
\end{figure}

To estimate the critical temperature, 
we have to extrapolate $\beta_{c}(N_t,N_s)$ to the thermodynamic limit 
and to the continuum limit. 
We perform the extrapolation to the thermodynamic limit 
using a finite-size scaling ansatz,
\begin{equation}
 \beta_{c}(N_t,N_s) = \beta_{c}(N_t,\infty) 
       - h \left( \frac{N_t}{\xi N_s}\right)^{3}.
\end{equation}
for first order phase transitions.
From the data for $\beta_{c}$ on anisotropic $12^3\times8$, $16^3\times8$ 
and $24^3\times8$ lattices with $\xi=2$, we find $h = 0.031(16)$ 
for $N_t/\xi=4$, 
as shown in Fig.~\ref{fig:finite_size_scaling}.
In a previous study on isotropic lattices, $h$ was found to be 
approximately independent of $N_t$ for $N_t=4$ and 6 \cite{qcdpax}. 
We extract $\beta_{c}(N_t,\infty)$ adopting $h = 0.031(16)$ 
for all $N_t$.

\begin{figure}[tb]
\begin{center}
\leavevmode
\epsfxsize=12cm\epsfbox{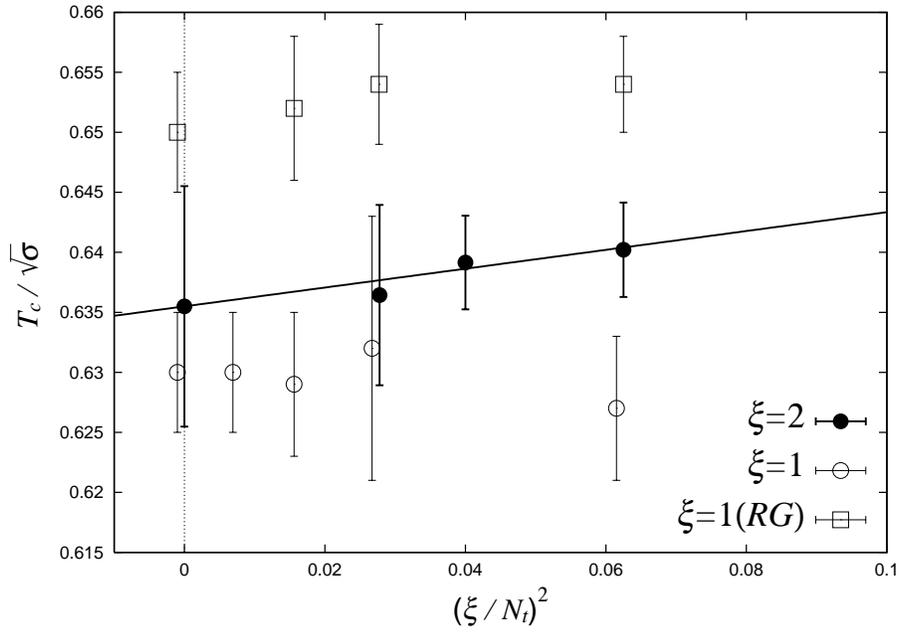}
\caption{Critical temperature $T_c / \sqrt{\sigma}$
         on isotropic and $\xi = 2$ anisotropic lattices.
         }
\label{fig:Tc}
\end{center}
\end{figure}

The critical temperature in units of the string tension is given by
\begin{equation}
\frac{T_c}{\sqrt{\sigma}} = 
\frac{\xi}
{N_t \left(a_s \sqrt{\sigma}\right)\left(\beta_c(N_t,\infty)\right)}
\end{equation}
using the fit result for Eq.~(\ref{allton_fitting_eq}).
The values of $T_{c} / \sqrt{\sigma}$ are 
summarized in Fig.~\ref{fig:Tc} and Table~\ref{tab:beta_c_lat}. 
The dominant part of the errors in $T_{c} / \sqrt{\sigma}$ is 
from the Allton fit for the string tension. 

Finally we extrapolate the results to the continuum limit 
assuming the leading $1/N_t^2$ scaling ansatz,
\begin{equation}
\left. F \right|_{N_t} 
= \left. F\right|_{\rm continuum} + \frac{c_F}{N_t^2}
\label{continuum-extrapolation}
\end{equation}
with $F = T_{c} / \sqrt{\sigma}$.
The extrapolation is shown in Fig.~\ref{fig:Tc}. 
In the continuum limit, we obtain 
\begin{equation}
 \frac{T_{c}}{\sqrt{\sigma}} = 0.635(10)
\label{eq:TcFinal}
\end{equation}
from the $\xi=2$ plaquette action. 

In Fig.~\ref{fig:Tc}, we also plot the results obtained on isotropic 
lattices using the plaquette action \cite{beinlich2} and the RG-improved 
action \cite{tsukuba,okamoto}. 
Our value of $T_c/\sqrt{\sigma}$ in the continuum limit is consistent 
with these results within the error of about 2\%. 
A more precise comparison would require the generation and analyses of 
potential data in a completely parallel manner, 
because, as discussed in \cite{okamoto}, 
numerical values of $T_c/\sqrt{\sigma}$ at a few percent level 
sensitively depend on the method used to determine the string tension.
We leave this issue for future studies.

\section{Pressure}
\label{sec:pressure}

\subsection{Integral method}

We use the integral method to calculate the pressure \cite{engels4}. 
This method is based on the relation $p = -f \equiv (T/V) \log Z(T,V)$ 
satisfied for a large homogeneous system, where $V=L_s^3$ is the spatial 
volume of the system in physical units and $Z$ is the partition function. 
Rewriting 
$\log Z = \int d\beta \frac{1}{Z} \frac{\partial Z}{\partial\beta}$, 
the pressure is given by 
\begin{equation}
\left. \frac{p}{T^{4}} \right|^{\beta}_{\beta_{0}} = 
\int_{\beta_{0}}^{\beta} d \beta^{\prime}
\Delta S(\beta^{\prime}),
\label{pressure}
\end{equation}
with
\begin{equation}
\Delta S (\beta) \equiv \xi \left(\frac{N_t}{\xi}\right)^{4}
                          \frac{1}{N_s^{3}N_t}
   \left. \frac{\partial \log Z}{\partial \beta} \right|_{\xi}.
\label{delta_S}
\end{equation}
For our anisotropic gauge action (\ref{lat-gauge-aniso}),
the derivative of $\log Z$ is given by
\begin{equation}
-\frac{\partial \log Z}{\partial \beta} 
= \bra \frac{S_{G}}{\beta} \ket
  + \beta \frac{\partial \xi_{0}(\beta,\xi)}{\partial \beta}
   \left( 
      \bra Q_t \ket 
      - \frac{\bra Q _s \ket}{\xi_{0}^{2}(\beta,\xi)}
   \right) 
  - (T=0 \,\, \mbox{contribution}). 
\end{equation}
We use symmetric $N_s^{3} \times \xi N_s$ lattices 
to calculate the $T = 0$ contribution.
For a sufficiently small $\beta_{0}$, $p(\beta_{0})$ can be neglected.

In order to keep the same accuracy of $\Delta S$ for the same physical 
lattice volume $L_s^3$ in units of the temperature $T$, 
the statistics of simulations should increase in proportion to 
$(\xi (N_t/\xi)^{4})^2 / (N_s^3 N_t) \propto N_t^4/\xi^3$. 
Here, the first factor arises from $\xi (N_t/\xi)^{4}$ in 
Eq.~(\ref{delta_S}), 
and the second factor $1/(N_s^3 N_t)$ from a suppression of fluctuations 
due to averaging over the lattice volume.
Taking into account the autocorrelation time which is proportional to 
$N_t^2$, the number of iterations should increase as $ \sim N_t^6$.

Integrating $\Delta S$ in $\beta$ using a cubic spline interpolation, 
we obtain the pressure. For the horizontal axis, 
we use the temperature in units of the critical temperature, 
\begin{equation}
 \frac{T}{T_{c}} 
 = \frac{(a_s \sqrt{\sigma})(\beta_{c})}
        {(a_s \sqrt{\sigma})(\beta)}.
\end{equation}
The errors from numerical integration are estimated by 
a jack-knife method in the following way \cite{okamoto}.
Since simulations at different $\beta$ are statistically independent, 
we sum up all the contributions from $\beta_i$ smaller than 
$\beta$ corresponding to the temperature $T$ by 
the naive error-propagation rule,
$\delta p(T) = \sqrt{\sum_i {\delta p_i(T)^2}}$, 
where $\delta_i p(T)$ at each simulation point $\beta_i$ is estimated 
by the jack-knife method.

\subsection{Finite spatial volume effects}

\begin{figure}[tb]
\begin{center}
\leavevmode
\epsfxsize=12cm\epsfbox{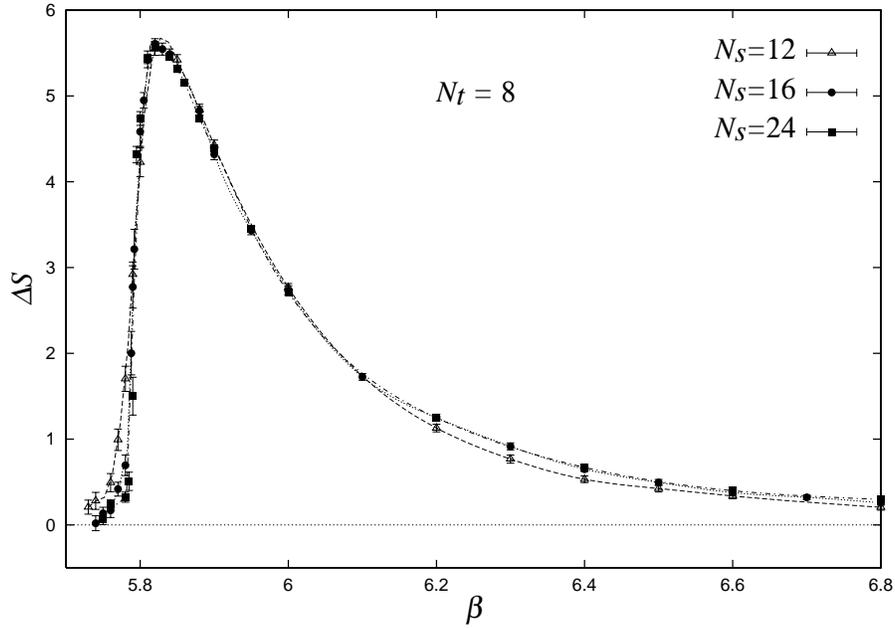}
\caption{Spatial lattice volume dependence in $\Delta S$ at $N_t/\xi = 4$ 
on $N_s=12$, 16 and 24 lattices with $\xi=2$.}
\label{fig-delta_S-12_24x8}
\end{center}
\end{figure}

\begin{figure}[tb]
\begin{center}
\leavevmode
\epsfxsize=12cm\epsfbox{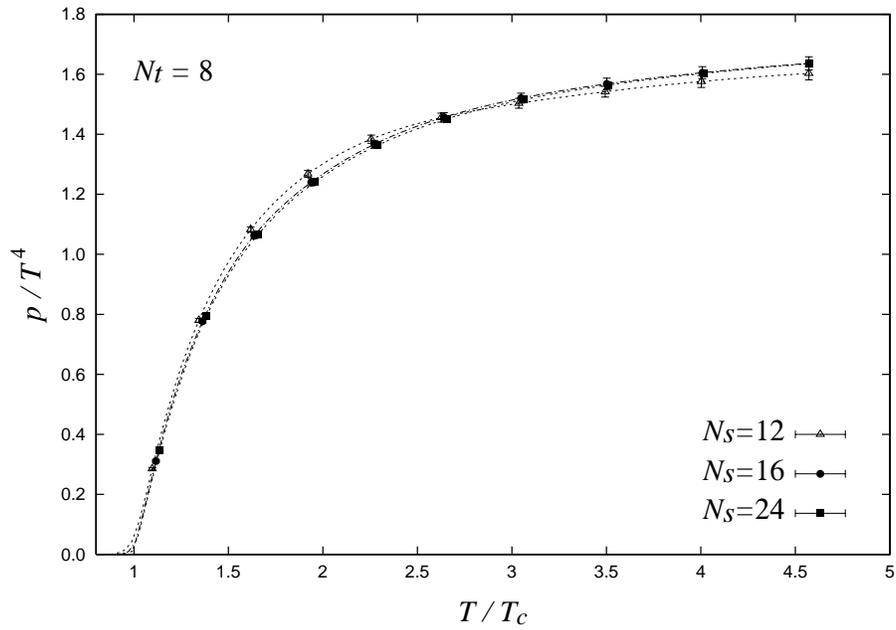}
\caption{Spatial volume dependence of the pressure $p/T^{4}$ 
on $\xi = 2$ anisotropic lattices with $N_t/\xi=4$.}
\label{fig-pressure-Nt8}
\end{center}
\end{figure}

\begin{figure}[tb]
\begin{center}
\leavevmode
\epsfxsize=12cm\epsfbox{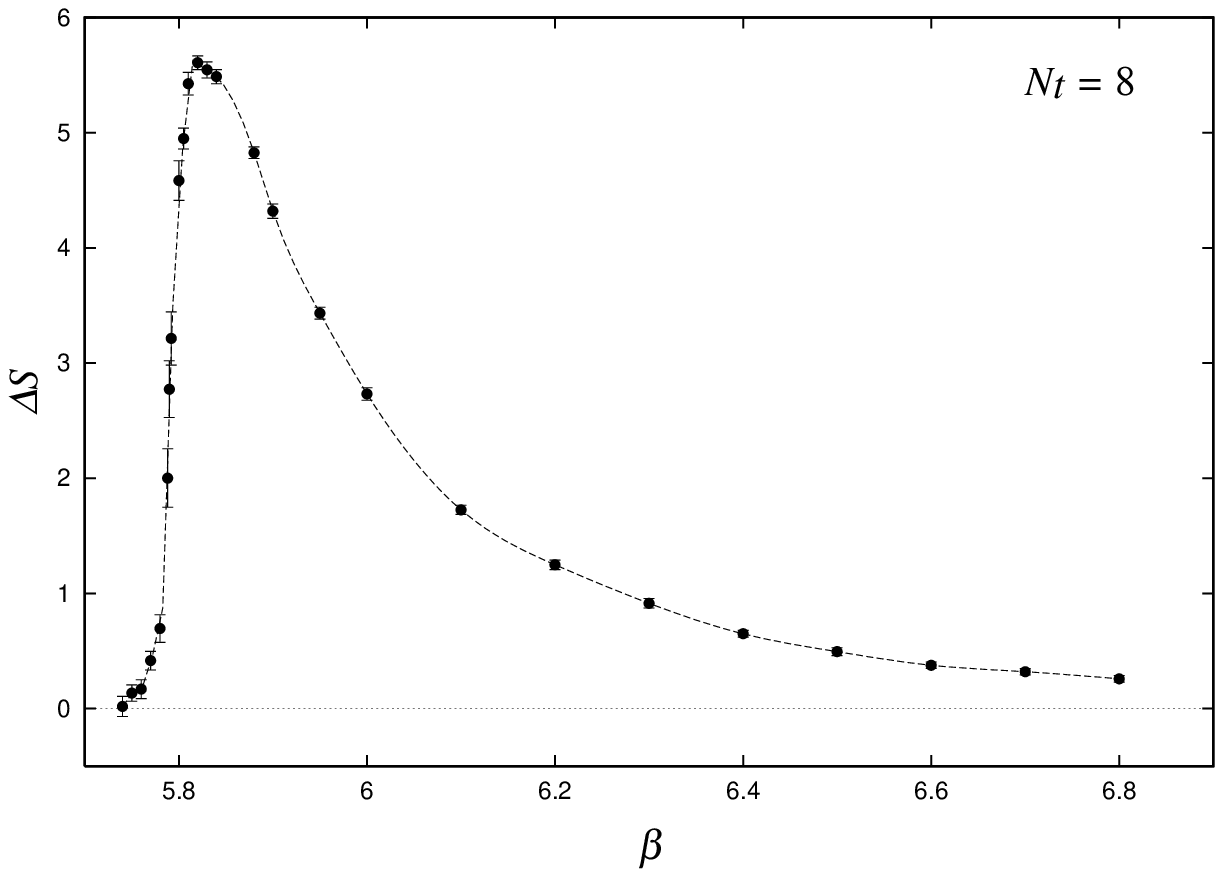}
\end{center}
\begin{center}
\leavevmode
\epsfxsize=12cm\epsfbox{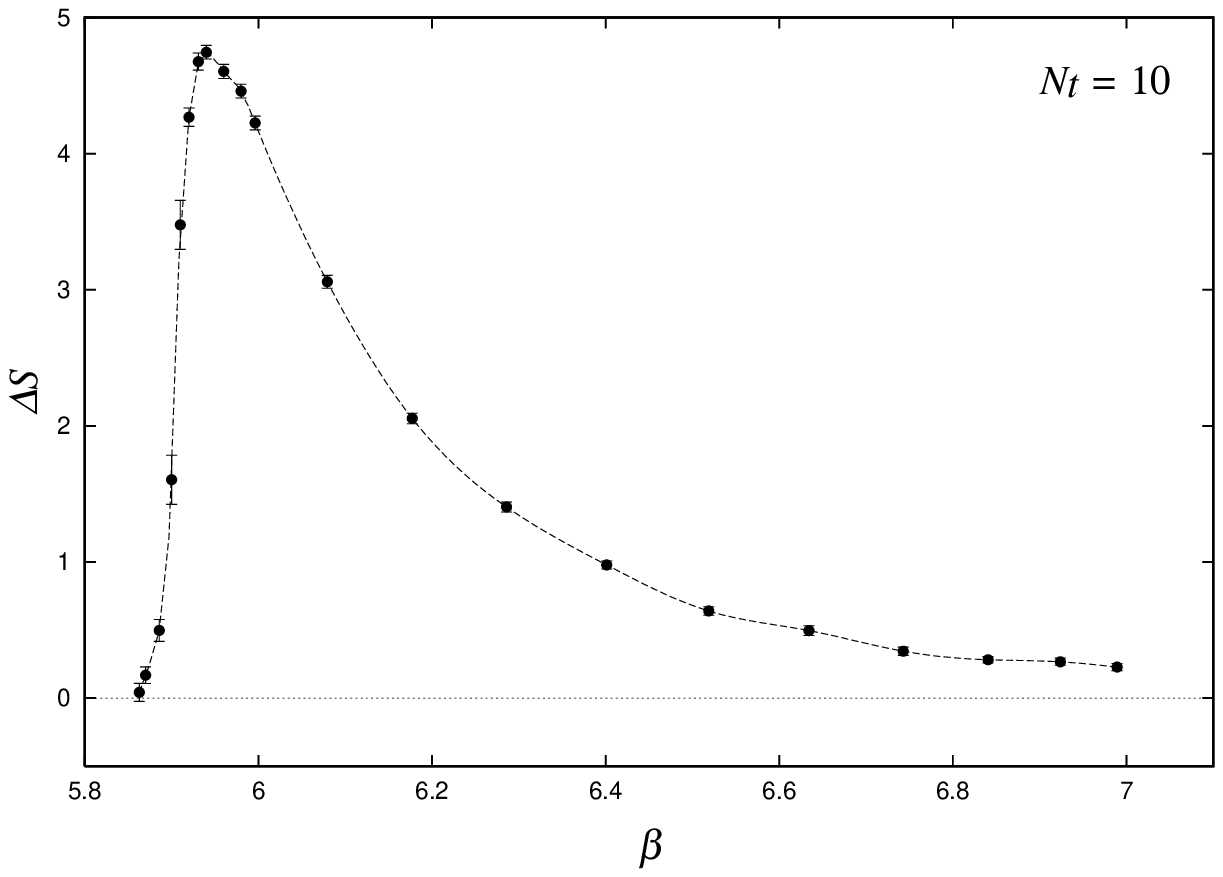}
\end{center}
\begin{center}
\leavevmode
\epsfxsize=12cm\epsfbox{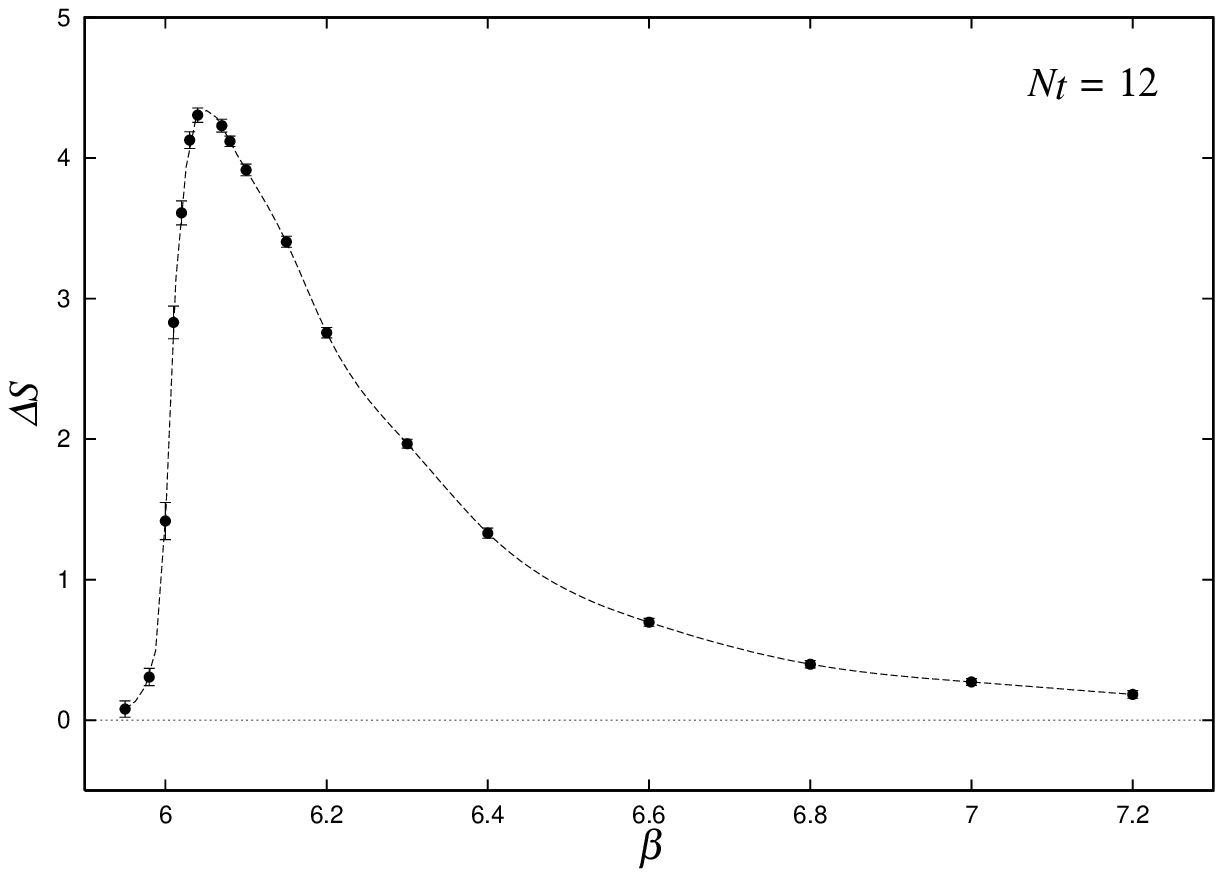}
\caption{$\Delta S$ on $N_t/\xi = 4$, 5 and 6 lattices with $\xi=2$.}
\label{fig-delta_S-8_12}
\end{center}
\end{figure}

\begin{figure}[tb]
\begin{center}
\leavevmode
\epsfxsize=12cm\epsfbox{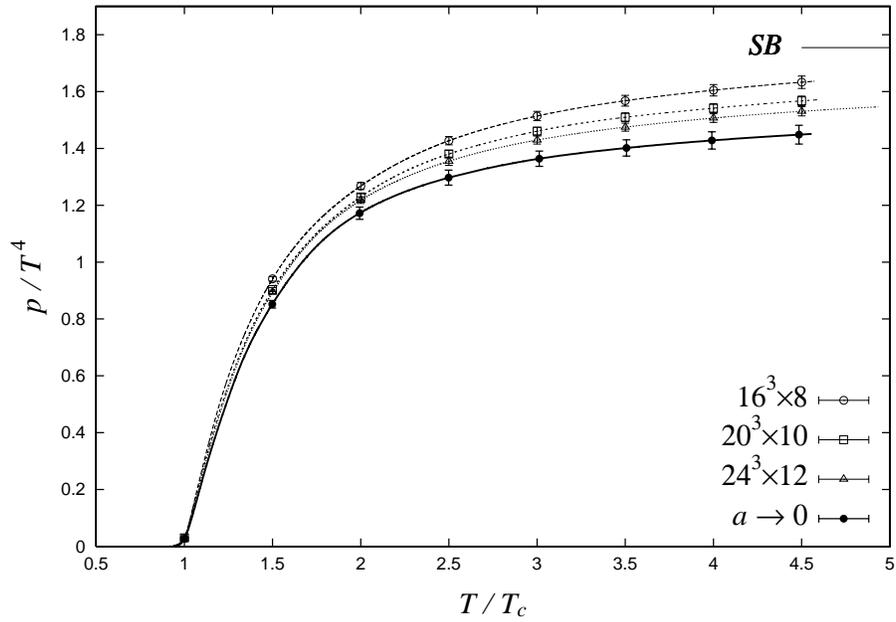}
\caption{Pressure $p/T^{4}$ on $\xi = 2$ anisotropic lattices.}
\label{fig-pressure}
\end{center}
\end{figure}

We first study the effects of finite spatial volume on the EOS. 
In Fig.~\ref{fig-delta_S-12_24x8}, we show the results for $\Delta S$ 
at $N_t/\xi=8/2$ with the aspect ratio $L_s T = N_s\xi/N_t = 3$, 4 and 6 
which correspond to $N_s=12$, 16 and 24, respectively.
Integrating $\Delta S$ in $\beta$, we obtain Fig.~\ref{fig-pressure-Nt8} 
for the pressure. 
We find that the data at $L_s T = 3$ is affected by sizable 
finite volume effects both at $T \sim T_c$ and at high temperatures. 
On the other hand, for the range of $T/T_c$ we study, 
the pressure does not change when the aspect ratio is increased from 
$L_s T=4$ to 6, indicating that the conventional choice $L_s T = 4$ is 
safe with the present precision of data. 
Hence, we choose $L_s T = 4$ for our studies of lattice spacing dependence.
Results for $\Delta S$ at $L_s T = 4$ with various $N_t$ 
are given in Fig.~\ref{fig-delta_S-8_12}.
Integrating the data using a cubic spline interpolation, as shown in 
the figures, we obtain the pressure plotted in Fig.~\ref{fig-pressure}.

\subsection{Continuum extrapolation}

\begin{figure}[tb]
\begin{center}
\leavevmode
\epsfxsize=12cm\epsfbox{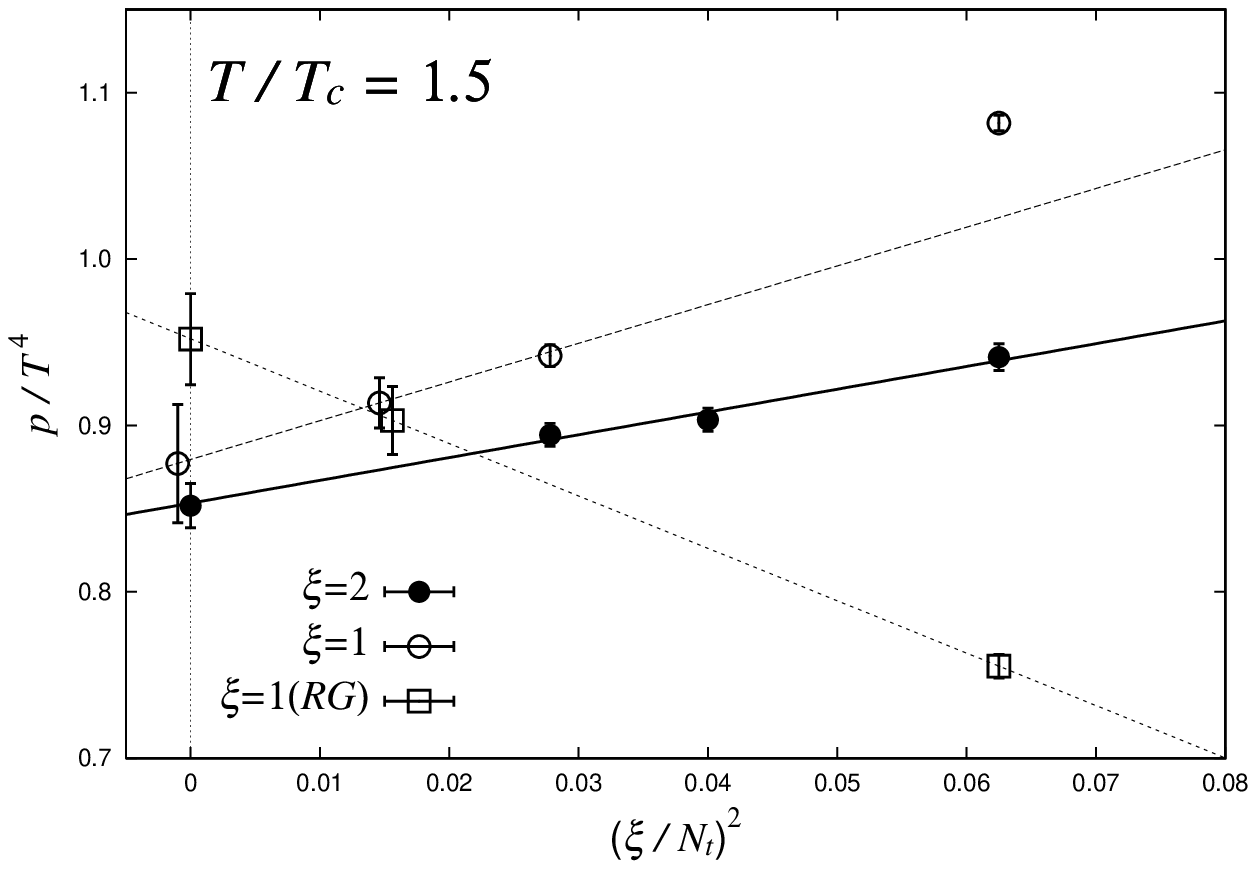}
\end{center}
\begin{center}
\leavevmode
\epsfxsize=12cm\epsfbox{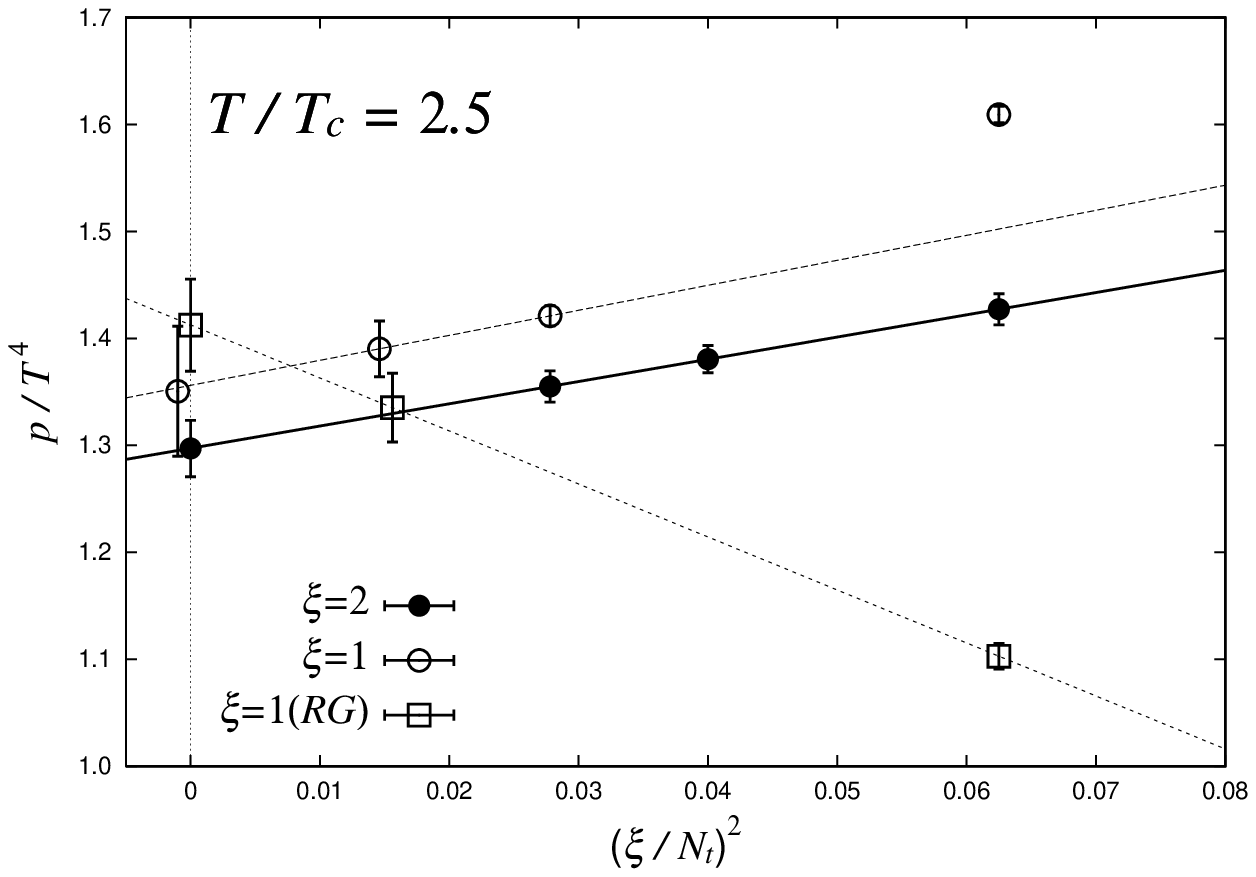}
\end{center}
\begin{center}
\leavevmode
\epsfxsize=12cm\epsfbox{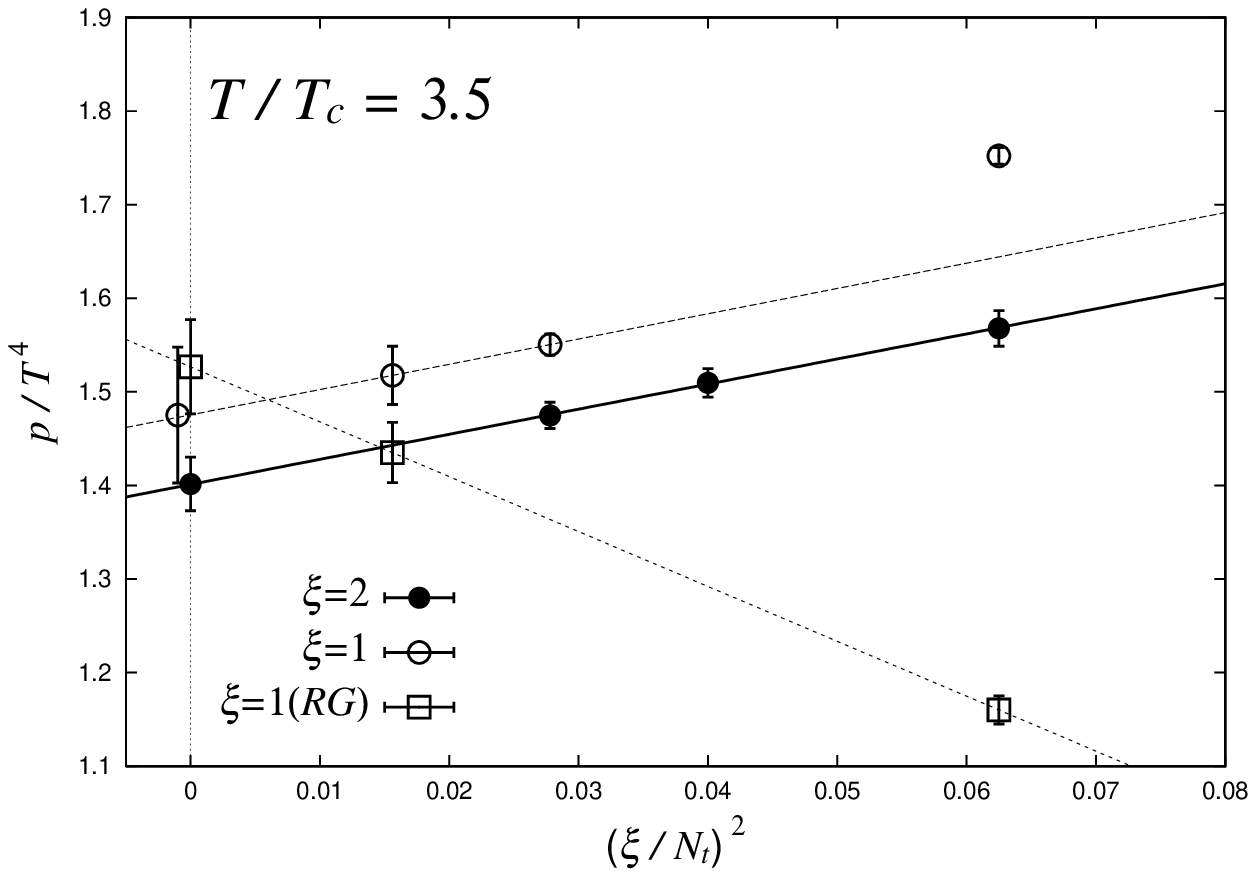}
\caption{Continuum extrapolation of the pressure $p/T^{4}$ 
at $T/T_c= 1.5$, 2.5 and 3.5.}
\label{fig-pressure-scaling}
\end{center}
\end{figure}

We now extrapolate the pressure to the continuum limit 
using the leading order ansatz of Eq.~(\ref{continuum-extrapolation}). 
Figure~\ref{fig-pressure-scaling} shows the pressure at $T/T_c=1.5$, 
2.5 and 3.5 as a function of $(\xi/N_t)^2$ (filled circles). 
For comparison, results from isotropic lattices using the 
plaquette action \cite{boyd} (open circles) 
and the RG-improved action \cite{okamoto} (open squares) 
are also plotted. 
For the $\xi=1$ plaquette data, we adopt the results of a reanalysis
made in Ref.\cite{okamoto} to commonly apply the scale from the Allton 
fit of the string tension and also the same error estimation method.

The advantage of using anisotropic lattices is apparent from 
Fig.~\ref{fig-pressure-scaling}.
On the coarsest lattice $N_t/\xi=4$, finite lattice 
spacing errors at $\xi=2$ are much smaller than those at $\xi=1$ 
with the same plaquette action. 
The pressure at $T = 2.5 T_{c}$, for example,  
on the isotropic $16^{3} \times 4$ lattice is 
larger than its continuum limit by about 20\%, 
while the deviation is only 5\% on the corresponding 
$16^{3} \times 8$ lattice with $\xi = 2$.
Furthermore, with the anisotropic $\xi=2$ data, 
the leading $1/N_t^2$ term describes the data well even 
at $N_t/\xi=4$ (the right-most point). 
Therefore, we can confidently perform an extrapolation 
to the continuum limit using three data points. 
In the case of the isotropic plaquette action, in contrast, 
the continuum extrapolation had to be made with 
only two data points at $N_t/\xi=6$ and 8.
In the continuum limit, our results for $\xi=2$ are slightly smaller 
than those from the isotropic plaquette action, but the results 
are consistent with each other within the error of about 5\% for 
the results from the isotropic action. 
It is worth observing that the $\xi=2$ results have smaller 
and more reliable errors of 2--3\%.  

  In order to quantitatively evaluate the benefit of anisotropic lattices, 
  we compare the computational cost to achieve comparable systematic 
  and statistical errors on isotropic and $\xi=2$ anisotropic lattices. 
  Choosing $T=2.5T_c$ as a typical example, we find that the deviation of 
  the pressure from the continuum limit ({\it i.e.}, the magnitude of 
  the systematic error due to finite lattice cutoffs) is comparable 
  between the isotropic $32^3 \times 8$ \cite{boyd} and our $\xi = 2$ 
  anisotropic $20^3 \times 10$ lattices,  
  {\it i.e.,} $p/T^4 = 1.390(26)$ on a $32^3\times 8$ lattice and 
  $p/T^4 = 1.381(13)$ on a $20^3\times 10$ lattice, both lattices having 
  the same spatial size $N_s a_s = 1.6/T_c$.
  The number of configurations to achieve these 
  statistical errors are 20,000--40,000 iterations for $\xi=1$
  and 50,000 for $\xi=2$, respectively.  
  Therefore, for the same statistical error, the relative computational 
  cost for a $\xi=2$ lattice over that for $\xi=1$ is conservatively 
  estimated as 
  $\left( (20^3 \times 10)\times 50000\right)/
   \left((32^3 \times 8)\times 4 \times 20000\right) \approx 1/5 $,  
  showing a factor 5 gain in the computational cost for the anisotropic 
  calculation in this example. 

In Fig.~\ref{fig-pressure-scaling} we also note that the results from 
the RG-improved action on isotropic lattices 
are higher by 7--10\% (about 2$\sigma$) than those from 
the present work in the continuum limit.
A possible origin of this discrepancy is the use of 
the $N_t/\xi=4$ data of the RG-improved action, which show a large 
(about 20\%) deviation from the continuum value. 
For a detailed test of consistency, we need more data points, 
say at $N_t/\xi=6$, from the RG-improved action. 

\begin{figure}[tb]
\begin{center}
\leavevmode
\epsfxsize=12cm\epsfbox{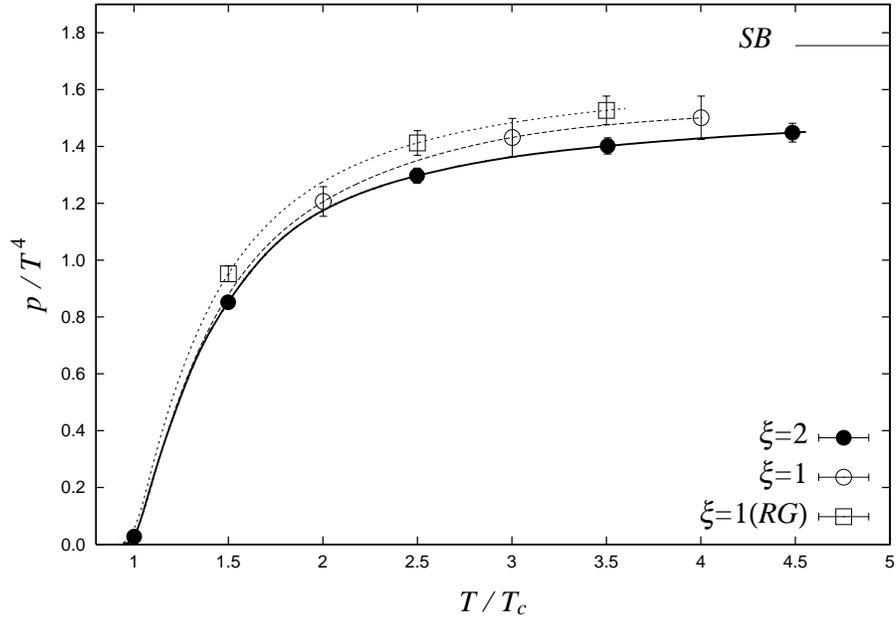}
\caption{Pressure $p/T^4$ in the continuum limit.}
\label{fig-pressure-continuum}
\end{center}
\end{figure}

Repeating the continuum extrapolation at other values of $T/T_c$, 
we obtain Fig.~\ref{fig-pressure-continuum}. 
Our results show a quite slow approach to the high temperature 
Stephan-Boltzmann limit, as reported also in previous studies 
on isotropic lattices \cite{boyd,okamoto}.

\section{Energy density}
\label{sec:energy}

We calculate the energy density $\epsilon$ by combining the 
results of $p/T^4$ with those for the interaction measure defined by 
\begin{equation}
\frac{\epsilon - 3 p}{T^{4}} 
= - a_s \left. \frac{\partial \beta}
                      {\partial a_s} \right|_{\xi} \Delta S.
\label{energy}
\end{equation}
The QCD beta function on anisotropic lattice 
$\left. \frac{\partial \beta}{\partial a_s} \right|_{\xi}$ is 
determined through the string tension $\sigma$ 
studied in Sec.~\ref{subsec:sigma}, 
\begin{equation}
 a_s \left. \frac{\partial \beta}{\partial a_s} \right|_{\xi}
 = \frac{12 b_{0}}{6 \left( b_{1}/b_{0} \right) \beta^{-1} -1}
   \,
   \frac{1 + c_{2} \hat{a}^{2} + c_{4} \hat{a}^{4}}
        {1 + 3 c_{2} \hat{a}^{2} + 5 c_{4} \hat{a}^{4}},
\end{equation}
where the coefficients $c_i$ are given in Eq.~(\ref{allton_fitting_eq}).
The error of the energy density is calculated 
by quadrature from the error of $3 p$ and that for 
$\epsilon-3p$, 
the latter being proportional to the error of $\Delta S$.

\begin{figure}[tb]
\begin{center}
\leavevmode
\epsfxsize=12cm\epsfbox{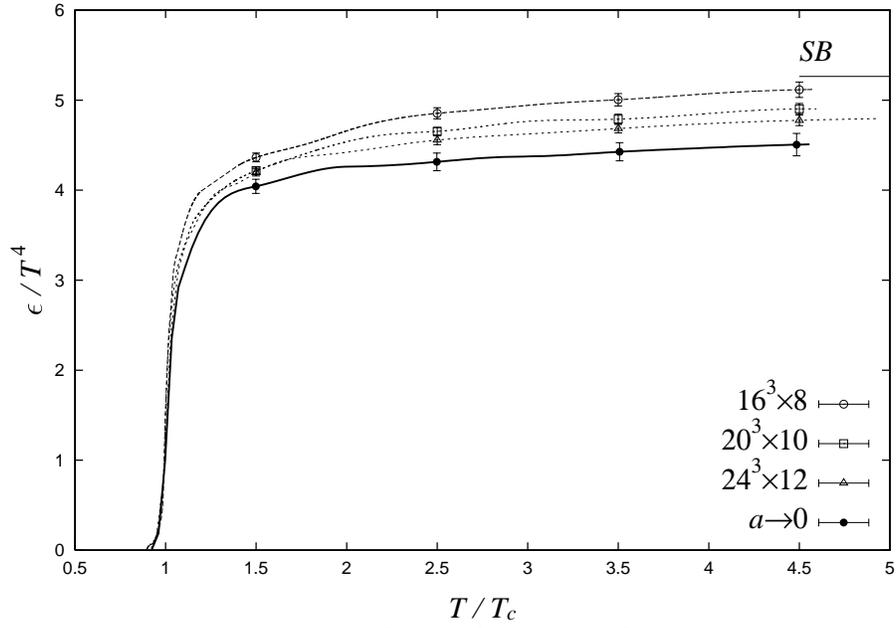}
\caption{$\epsilon/T^{4}$ 
         on anisotropic $16^{3} \times 8$, $20^{3} \times 10$ 
         and $24^{3} \times 12$ lattices with $\xi = 2$.} 
\label{fig-energy}
\end{center}
\end{figure}

\begin{figure}[tb]
\begin{center}
\leavevmode
\epsfxsize=12cm\epsfbox{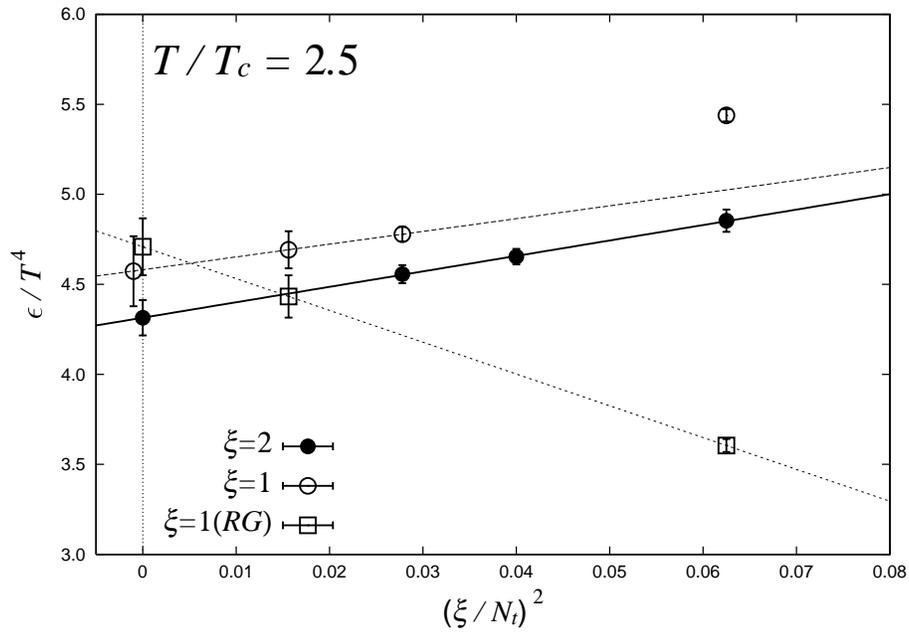}
\caption{Continuum extrapolation of the energy density 
$\epsilon/T^{4}$ at $T = 2.5 T_{c}$.} 
\label{fig-energy-scaling}
\end{center}
\end{figure}

\begin{figure}[tb]
\begin{center}
\leavevmode
\epsfxsize=12cm\epsfbox{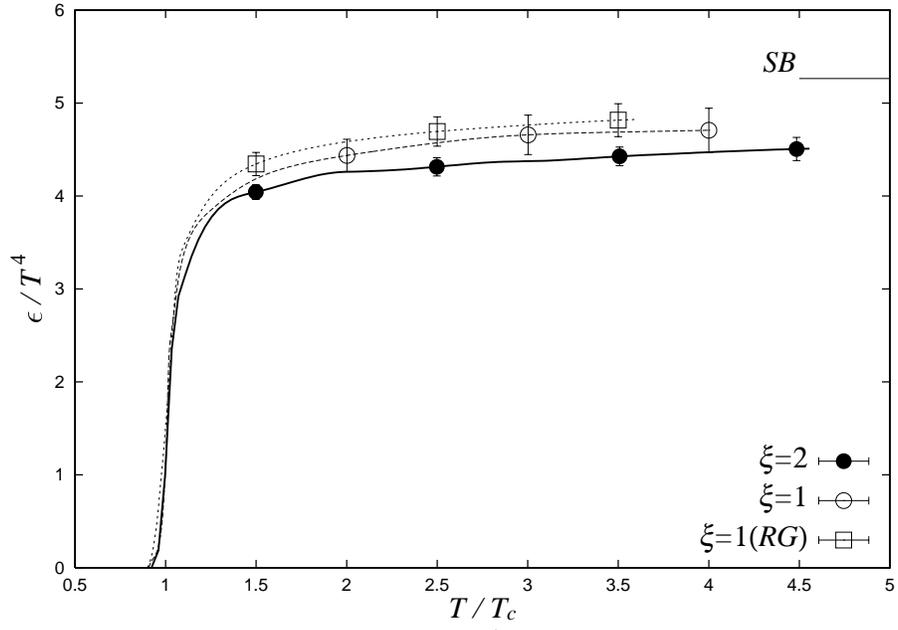}
\caption{Energy density $\epsilon/T^{4}$ in the continuum limit.} 
\label{fig-energy-continuum}
\end{center}
\end{figure}

The results for the energy density are shown in 
Figs.~\ref{fig-energy} and \ref{fig-energy-scaling}.
As in the case of the pressure  
the leading scaling behavior is well followed by our $\xi=2$ 
data from $N_t/\xi = 4$, which allows us to extrapolate to the 
continuum limit reliably. 
The results for the energy density in the continuum limit are 
compared with the previous results in Fig.~\ref{fig-energy-continuum}. 
Our $\xi=2$ plaquette action leads
to an energy density which is slightly smaller than, 
but consistent with that from the $\xi=1$ plaquette action, 
but is about 7--10\% (about 2$\sigma$) smaller than that from the 
$\xi=1$ RG action.
More work is required to clarify the origin of the small 
discrepancy with the RG action.

\section{Conclusion}
\label{sec:summary}

We have studied the continuum limit of the equation of state 
in SU(3) gauge theory on anisotropic lattices 
with the anisotropy $\xi \equiv a_s/a_t =2$, 
using the standard plaquette gauge action. 
Anisotropic lattices are shown to be more efficient 
in calculating thermodynamic quantities than isotropic lattices. 
We found that the cutoff errors in the pressure and energy density 
are much smaller than corresponding isotropic lattice 
results at small values of $N_t/\xi$.
  The computational cost for $\xi=2$ lattices is about 1/5 of that 
  for $\xi=1$ lattices.
We also found that the leading scaling behavior is well satisfied 
already from $N_t/\xi = 4$, 
which enabled us to perform continuum extrapolations 
with three data points at $N_t/\xi=4$, 5 and 6. 
The equation of state in the continuum limit agrees with that obtained 
on isotropic lattice using the same action,
but have much smaller and better controlled errors. 
The benefit of anisotropic lattice demonstrated here  will be indispensable 
for extraction of continuum predictions for the equation of state,
when we include dynamical quarks.

\section*{Acknowledgements}
This work is supported in part by Grants-in-Aid of the Ministry of Education 
(Nos.~10640246, 
10640248, 
11640250, 
11640294, 
12014202, 
12304011, 
12640253, 
12740133, 
13640260  
). 
SE and M. Okamoto are JSPS Research Fellows. 
VL is supported by the Research for Future Program of JSPS
(No. JSPS-RFTF 97P01102).
Simulations were performed on the parallel computer CP-PACS 
at the Center for Computational Physics, University of Tsukuba.



\newpage

\begin{table}[tb]
\begin{center}
\begin{tabular}{llrr} 
 lattice            &  $\beta$     & bin size & \# of iter. \\
\hline
 $12^{3} \times 8$   & 5.73--6.80  &  1600    &  40 000 \\
 $16^{3} \times 8*$  & 5.74--6.80  &   800    &  20 000 \\
 $24^{3} \times 8$   & 5.75--6.80  &   400    &  10 000 \\
 $20^{3} \times 10*$ & 5.86--6.98  &  2000    &  50 000 \\
 $24^{3} \times 12*$ & 5.95--7.20  &  4000    & 100 000 \\ 
\hline
 $12^{3} \times 24$  & 5.74--6.80  &   400    &  10 000 \\
 $16^{3} \times 32*$ & 5.74--6.80  &   200    &   5 000 \\
 $20^{3} \times 40*$ & 5.86--6.98  &   500    &  12 500 \\
 $24^{3} \times 48$  & 5.75--5.90  &   100    &   2 500 \\
 $24^{3} \times 48*$ & 5.95--7.20  &  1000    &  25 000
\end{tabular}
\caption{Simulation parameters. 
	 Main runs are marked by star ($*$).}
\label{tab:simulation_parameters}
\end{center}
\end{table}

\begin{table}[tb]
\begin{center}
\begin{tabular}{llrr} 
 lattice             &  $\beta$     & bin size & \# of iter. \\
\hline
 $12^{3} \times 8$   & 5.790, 5.791 &  8000    &  80 000 \\
 $16^{3} \times 8$   & 5.790, 5.792 &  4000    &  40 000 \\
 $24^{3} \times 8$   & 5.791, 5.792 &  4000    &  40 000 \\
 $20^{3} \times 10$  & 5.903, 5.907 &  5000    &  50 000 \\
 $24^{3} \times 12$  & 6.004, 6.006 & 10000    & 100 000 \\
\end{tabular}
\caption{Simulation parameters for determination of critical couplings.}
\label{tab:simulation_parameters-critical_coupling}
\end{center}
\end{table}

\begin{table}[tb]
\begin{center}
\begin{tabular}{llllll}
&&\multicolumn{2}{c}{$16^{3} \times 8$}
 &\multicolumn{2}{c}{$16^{3} \times 32$} \\ 
\hline
 $\beta$ & $\xi_{0}$ & $P_{ss}$ & $P_{st}$
                     & $P_{ss}$ & $P_{st}$ 
\\ 
\hline
 5.740 & 1.66279318 & 0.448467(31) & 0.679985(12)
                    & 0.448490(28) & 0.679979(11) \\
 5.750 & 1.66473308 & 0.450693(24) & 0.681412(11)
                    & 0.450641(21) & 0.681384(8) \\
 5.760 & 1.66664410 & 0.452784(33) & 0.682783(13)
                    & 0.452731(22) & 0.682747(9) \\
 5.770 & 1.66852693 & 0.454935(29) & 0.684175(13)
                    & 0.454758(24) & 0.684090(9) \\
 5.780 & 1.67038223 & 0.457024(53) & 0.685533(22)
                    & 0.456720(21) & 0.685392(8) \\
 5.788 & 1.67184708 & 0.459186(116) & 0.686823(49)
                    & 0.458272(30) & 0.686419(11) \\
 5.790 & 1.67221065 & 0.459930(109) & 0.687240(48)
                    & 0.458678(26) & 0.686679(11) \\
 5.792 & 1.67257316 & 0.460517(104) & 0.687578(45)
                    & 0.459056(22) & 0.686929(9) \\
 5.800 & 1.67401280 & 0.462698(75) & 0.688873(33)
                    & 0.460586(22) & 0.687949(9) \\
 5.805 & 1.67490422 & 0.463825(34) & 0.689587(15)
                    & 0.461565(21) & 0.688588(9) \\
 5.810 & 1.67578929 & 0.464912(40) & 0.690278(17)
                    & 0.462446(20) & 0.689181(9) \\
 5.820 & 1.67754071 & 0.466746(21) & 0.691520(10)
                    & 0.464241(17) & 0.690383(6) \\
 5.830 & 1.67926762 & 0.468486(24) & 0.692704(10)
                    & 0.466022(21) & 0.691578(9) \\
 5.840 & 1.68097058 & 0.470122(18) & 0.693839(8)
                    & 0.467707(24) & 0.692722(9) \\
 5.880 & 1.68755324 & 0.476195(15) & 0.698142(7)
                    & 0.474205(17) & 0.697145(7) \\
 5.900 & 1.69071395 & 0.478994(18) & 0.700156(9)
                    & 0.477282(22) & 0.699255(9) \\
 5.950 & 1.69826359 & 0.485606(15) & 0.704933(7)
                    & 0.484390(18) & 0.704199(7) \\
 6.000 & 1.70535029 & 0.491774(15) & 0.709406(6)
                    & 0.490955(20) & 0.708801(9) \\
 6.100 & 1.71830738 & 0.503237(14) & 0.717652(6)
                    & 0.502986(14) & 0.717230(5) \\
 6.200 & 1.72987892 & 0.513833(11) & 0.725175(6)
                    & 0.513839(14) & 0.724837(5) \\
 6.300 & 1.74029271 & 0.523743(10) & 0.732106(4)
                    & 0.523915(15) & 0.731827(7) \\
 6.400 & 1.74972820 & 0.533075(11) & 0.738552(4)
                    & 0.533401(9) & 0.738316(3) \\
 6.500 & 1.75832876 & 0.541970(13) & 0.744586(5)
                    & 0.542362(8) & 0.744378(5) \\
 6.600 & 1.76621035 & 0.550391(8) & 0.750250(3)
                    & 0.550854(10) & 0.750058(4) \\
 6.700 & 1.77346785 & 0.558485(9) & 0.755608(4)
                    & 0.558959(9) & 0.755427(4) \\
 6.800 & 1.78017964 & 0.566215(12) & 0.760672(4)
                    & 0.566716(8) & 0.760501(4) \\
\end{tabular}
\caption{
Plaquette expectation values 
on $16^{3} \times 8$ and $16^{3} \times 32$ lattices
with $\xi=2$.
}
\label{tab:MC_results-16x8_32}
\end{center}
\end{table}

\begin{table}[tb]
\begin{center}
\begin{tabular}{llllll}
&&\multicolumn{2}{c}{$20^{3} \times 10$}
 &\multicolumn{2}{c}{$20^{3} \times 40$} \\ 
\hline
 $\beta$ & $\xi_{0}$ & $P_{ss}$ & $P_{st}$
                     & $P_{ss}$ & $P_{st}$
\\ 
\hline
 5.86288916 & 1.68478116 & 0.4715286(90) & 0.6953072(38)
                    & 0.4715194(98) & 0.6953039(38) \\
 5.87 & 1.68594094 & 0.4726803(97) & 0.6960907(37)
                    & 0.4726453(79) & 0.6960771(33) \\
 5.88583578 & 1.68848420 & 0.4752043(113) & 0.6978062(52)
                    & 0.4751072(93) & 0.6977655(41) \\
 5.90 & 1.69071395 & 0.4775533(342) & 0.6993698(144)
                    & 0.4772612(79) & 0.6992430(33) \\
 5.91 & 1.69226327 & 0.4793349(340) & 0.7005240(144)
                    & 0.4787235(65) & 0.7002573(30) \\
 5.92 & 1.69379248 & 0.4809915(113) & 0.7016191(50)
                    & 0.4801832(57) & 0.7012665(26) \\
 5.93084722 & 1.69542899 & 0.4826008(89) & 0.7027227(39)
                    & 0.4817182(78) & 0.7023359(35) \\
 5.94 & 1.69679224 & 0.4838962(61) & 0.7036250(30)
                    & 0.4830113(60) & 0.7032314(30) \\
 5.96 & 1.69971645 & 0.4865820(62) & 0.7055225(30)
                    & 0.4857427(62) & 0.7051382(32) \\
 5.98 & 1.70256818 & 0.4891795(54) & 0.7073650(25)
                    & 0.4883883(83) & 0.7069900(34) \\
 5.9961937 & 1.70482605 & 0.4912217(55) & 0.7088160(30)
                    & 0.4904832(71) & 0.7084591(30) \\
 6.0793640 & 1.71575557 & 0.5010417(44) & 0.7158270(31)
                    & 0.5005840(62) & 0.7155576(27) \\
 6.17716193 & 1.72734556 & 0.5116532(54) & 0.7233550(25)
                    & 0.5114357(43) & 0.7231598(22) \\
 6.28582916 & 1.73888020 & 0.5225991(56) & 0.7310157(21)
                    & 0.5225280(53) & 0.7308687(21) \\
 6.40118969 & 1.74983517 & 0.5334631(32) & 0.7385009(19)
                    & 0.5334926(43) & 0.7383839(17) \\
 6.51881026 & 1.75986308 & 0.5438681(48) & 0.7455581(19)
                    & 0.5439702(40) & 0.7454657(19) \\
 6.63417079 & 1.76875624 & 0.5535144(38) & 0.7520032(19)
                    & 0.5536476(51) & 0.7519204(23) \\
 6.74283803 & 1.77640579 & 0.5621461(45) & 0.7576970(23)
                    & 0.5623098(36) & 0.7576251(14) \\
 6.84063596 & 1.78276647 & 0.5695876(32) & 0.7625475(17)
                    & 0.5697626(34) & 0.7624799(11) \\
 6.92380626 & 1.78783002 & 0.5756793(33) & 0.7664882(18)
                    & 0.5758587(31) & 0.7664206(16) \\
 6.98915275 & 1.79160648 & 0.5803248(35) & 0.7694702(14)
                    & 0.5805094(41) & 0.7694057(18) \\
\end{tabular}
\caption{
Plaquette expectation values 
on $20^{3} \times 10$ and $20^{3} \times 40$ lattices with $\xi=2$.}
\label{tab:MC_results-20x10_40}
\end{center}
\end{table}

\begin{table}[tb]
\begin{center}
\begin{tabular}{llllll}
&&\multicolumn{2}{c}{$24^{3} \times 12$}
 &\multicolumn{2}{c}{$24^{3} \times 48$} 
\\ 
\hline
 $\beta$ & $\xi_{0}$ & $P_{ss}$ & $P_{st}$
                     & $P_{ss}$ & $P_{st}$
\\ 
\hline
 5.95 & 1.69826359 & 0.4843851(27) & 0.7041916(13)
                    & 0.4843789(45) & 0.7041883(19) \\
 5.98 & 1.70256818 & 0.4884099(39) & 0.7070003(19)
                    & 0.4883825(35) & 0.7069880(15) \\
 6.00 & 1.70535029 & 0.4911005(118) & 0.7088537(50)
                    & 0.4909663(38) & 0.7087977(14) \\
 6.01 & 1.70671610 & 0.4924924(104) & 0.7097962(43)
                    & 0.4922291(37) & 0.7096838(15) \\
 6.02 & 1.70806552 & 0.4938053(64) & 0.7107011(32)
                    & 0.4934718(30) & 0.7105575(13) \\
 6.03 & 1.70939887 & 0.4950807(40) & 0.7115881(16)
                    & 0.4947043(36) & 0.7114232(17) \\
 6.04 & 1.71071646 & 0.4963132(30) & 0.7124510(16)
                    & 0.4959199(32) & 0.7122791(13) \\
 6.07 & 1.71457763 & 0.4998634(27) & 0.7149595(10)
                    & 0.4994891(31) & 0.7147889(15) \\
 6.08 & 1.71583512 & 0.5010194(19) & 0.7157747(6)
                    & 0.5006575(31) & 0.7156082(13) \\
 6.10 & 1.71830738 & 0.5032879(22) & 0.7173807(10)
                    & 0.5029551(29) & 0.7172208(13) \\
 6.15 & 1.72425080 & 0.5087787(26) & 0.7212576(10)
                    & 0.5085106(19) & 0.7211154(12) \\
 6.20 & 1.72987892 & 0.5140368(26) & 0.7249549(12)
                    & 0.5138372(20) & 0.7248368(8) \\ 
 6.30 & 1.74029271 & 0.5240287(21) & 0.7319188(8)
                    & 0.5239220(23) & 0.7318284(10) \\
 6.40 & 1.74972820 & 0.5334259(25) & 0.7383798(11)
                    & 0.5333873(23) & 0.7383125(9) \\
 6.60 & 1.76621035 & 0.5508062(15) & 0.7501014(7)
                    & 0.5508372(22) & 0.7500563(9) \\
 6.80 & 1.78017964 & 0.5666348(15) & 0.7605281(6)
                    & 0.5667010(21) & 0.7604924(9) \\
 7.00 & 1.79221720 & 0.5811933(20) & 0.7699251(8)
                    & 0.5812721(12) & 0.7698933(6) \\
 7.20 & 1.80273290 & 0.5946688(17) & 0.7784726(9)
                    & 0.5947568(18) & 0.7784435(8) \\
\end{tabular}
\caption{
Plaquette expectation values
on $24^{3} \times 12$ and $24^{3} \times 48$ lattices with $\xi=2$.}
\label{tab:MC_results-24x12_48}
\end{center}
\end{table}


\begin{table}[tb]
\begin{center}
\begin{tabular}{llrr}
$\beta$ & lattice            & $N_{opt}$ & \# of conf. \\ 
\hline
5.7     & $16^{3} \times 32$ &    3    &      800   \\
5.8     & $16^{3} \times 32$ &    5    &      800   \\
5.9     & $16^{3} \times 32$ &    6    &      800   \\
6.0     & $16^{3} \times 32$ &    8    &      600   \\
        & $24^{3} \times 48$ &    8    &      100   \\
6.1     & $16^{3} \times 32$ &   10    &      400   \\
6.3     & $16^{3} \times 32$ &   16    &      300   \\
        & $24^{3} \times 48$ &   20    &      100   \\
6.5     & $24^{3} \times 48$ &   30    &      100 
\end{tabular}
\caption{Simulation parameters for static quark potential at zero 
temperature.}
\label{tab:pot_simulations}
\end{center}
\end{table}

\begin{table}[tb]
\begin{center}
\begin{tabular}{lllllllllr}
$\beta$ & lattice
 &$a_s\sqrt{\sigma}$ &$L_{s}$[fm] &$\hat{T}$ &$\hat{R}_{min}$ &$V_{0}$  
        &     $e$    &  $l$     & $\chi^2 / N_{DF}$
\\ \hline
5.7    & $16^{3} \times 32*$
       & 0.4794(66) & 3.49 & 5         & $\sqrt{5}$  & 0.677(36)
       & 0.305(50)  & 0.934(122)       &     5.81   
\\ 
5.8    & $16^{3} \times 32*$
       & 0.3804(24) & 2.77 & 6         & $\sqrt{5}$  & 0.720(11)
       & 0.326(16)  & 0.647(49)        &     3.07   
\\ 
5.9    & $16^{3} \times 32*$
       & 0.3190(18) & 2.32 & 7         & $\sqrt{5}$  & 0.688(7)
       & 0.284(11)  & 0.501(43)        &     3.20
\\
6.0    & $16^{3} \times 32$
       & 0.2667(21) & 1.94 & 8         & $\sqrt{6}$  & 0.685(8)
       & 0.283(14)  & 0.396(73)        &     0.93   
\\
       & $24^{3} \times 48*$
       & 0.2611(31) & 2.85 & 8         & $\sqrt{6}$  & 0.699(11)
       & 0.310(19)  & 0.565(82)        &     2.05   
\\ 
6.1    & $16^{3} \times 32*$
       & 0.2224(20) & 1.61 & 8         &  2$\sqrt{2}$& 0.686(6)
       & 0.297(13)  & 0.375(61)        &     1.97   
\\
6.3    & $16^{3} \times 32$
       & 0.1656(19) & 1.20 & 9         &  $\sqrt{6}$ & 0.653(5)
       & 0.281(9)   & 0.239(67)        &     0.95    
\\
       & $24^{3} \times 48*$
       & 0.1661(20) & 1.81 & 9         & $\sqrt{6}$  & 0.657(5)
       & 0.294(9)   & 0.323(68)         &    1.72    
\\ 
6.5    & $24^{3} \times 48*$
       & 0.1242(21) & 1.35 & 9         & $\sqrt{6}$  & 0.622(3)
       & 0.279(6)   & 0.247(47)        &    1.75    
\end{tabular}
\caption{Results for the potential parameters 
	on $\xi = 2$ anisotropic lattices with the plaquette action.
        The spatial lattice size $L_s$ is computed using 
	$\sqrt{\sigma} = 440$ MeV.}
\label{tab:string_beta}
\end{center}
\end{table}

\begin{table}[tb]
\begin{center}
\begin{tabular}{llllll}
 $N_s^{3} \times N_t$   
 & $12^{3} \times 8$      
 & $16^{3} \times 8$      
 & $24^{3} \times 8$ 
 & $20^{3} \times 10$    & $24^{3} \times 12$
\\ \hline
 $\beta_{c}(N_t,N_s)$     
 & 5.79037(40)            
 & 5.79081(54)            
 & 5.79138(31)
 & 5.90494(92)           & 6.00464(67)
\\ \hline
 $\beta_{c}(N_t,\infty)$
 & & 5.79149(34)   &         & 5.90543(116)          & 6.00512(91)
\\ \hline
 $T_{c} / \sqrt{\sigma}$
 & & 0.6402(39)    &         & 0.6392(39)            & 0.6364(75)
\end{tabular}
\label{tab:beta_c_lat}
\caption{Critical coupling and temperature on anisotropic $\xi = 2$ 
	lattices. Results for $T_{c} / \sqrt{\sigma}$ are obtained 
	in the thermodynamic limit.}
\end{center}
\end{table}

\newpage

\end{document}